\documentclass[letterpaper]{JHEP3} 
\usepackage{subfigure}
\usepackage{graphicx, epsfig,psfrag}

\newcommand{\be}{\begin{equation}}
\newcommand{\ee}{\end{equation}}
\newcommand{\bea}{\begin{eqnarray}}
\newcommand{\eea}{\end{eqnarray}}

\def\etmiss{E\!\!\!\!\slash_{T}}
\def\ptmiss{p\!\!\!\slash_{T}}
\def\dm{\Delta M_{TA} }
\def\tt{t\bar t}
\def\gev{\rm GeV}
\def\tev{\rm TeV}
\def\fbi{{\rm fb}^{-1}}

\preprint{MADPH-08-1504,~
NSF-KITP-08-27,~
FERMILAB-PUB-08-059-T,~
UCB-PTH-08/05,~
PUPT-2260}

\title{\Large Top-quark Pair plus Large Missing Energy at the LHC}
 
\author{Tao Han$^{1,2}$, Rakhi Mahbubani$^{2,3}$, Devin G. E. Walker$^{2,4}$,
  Lian-Tao Wang$^{2,5}$ \\ 
$^1${\it Department of Physics, University of Wisconsin, Madison, WI 53706} \\
$^2$ {KITP, University of California, Santa Barbara, CA 93107} \\ 
$^3${\it  Fermi National Acceleration Laboratory, Batavia, IL 60510} \\
$^4${\it Department of Physics, University of California, Berkeley, CA 94720 and
Theoretical Physics Group, Lawrence Berkeley National Laboratory,
Berkeley, CA 94720}\\
$^4${\it Department of Physics, Princeton University, Princeton, NJ. 08544}}
\date{\today}

\abstract{
We study methods of extracting new physics signals in final states with
a top-quark pair plus large missing energy at the LHC. 
We consider two typical examples of such new
physics:  pair production of a fermionic top partner (a $T'$ in Little Higgs 
models for example) and of a scalar top partner (a $\tilde{t}$ in SUSY).
With a commonly-adopted discrete symmetry under which non Standard Model 
particles are odd, the top partner is assumed to decay 
predominantly to a top quark plus a massive neutral stable particle $A^0$.
We focus on the case in which one of the top quarks decays
leptonically and the other decays hadronically, $pp \to {\tt} A^0A^0 X
\to bj_1j_2\ \bar b\ell^- \bar\nu \ A^0A^0\ X + c.c.$, 
where the $A^0$s escape detection.  We identify a key parameter for the signal 
observation:  the mass splitting between the top partner and the missing particle. 
We reconstruct a transverse mass for the lepton-missing
transverse energy system to separate the real $W$ background from the signal 
and propose a definition for the
reconstructed top quark mass that allows it to take unphysical values as an
indication of new physics. 
We perform a scan over the two masses to map out the discovery reach at the
LHC in this channel. We also comment on the possibility of
distinguishing between scalar and fermionic top partners using collider
signatures.}

\begin{document}
\section{Introduction}

Although the Standard Model (SM) of elementary  particle physics 
provides a very successful description of existing experiments 
at the highest energies currently accessible at colliders, it is
anticipated that new physics will show up at the unexplored TeV-scale
territory.  
High energy physics will thus experience the excitement of major discoveries 
in the next few years when the CERN  Large Hadron Collider (LHC) opens up 
the new energy frontier. 
In addition to the long-awaited higgs boson, the particle responsible 
for the generation of mass, from naturalness arguments we hope to see a 
glimpse of some new physics at the LHC. Examples of popular scenarios of 
new physics include the Minimal Supersymmetric 
Standard Model (MSSM) \cite{Dimopoulos:1981zb} 
and its variants \cite{Cohen:1996vb,Kraml:2006ga}; models of
new strong dynamics \cite{technicolor,topcolor,topseesaw,tcreview,top-composite}  
or a composite Higgs at the TeV scale \cite{Kaplan:1983sm}; Little Higgs 
theories \cite{LHreview} and electroweak-scale extra dimensions
\cite{Arkani-Hamed:1998rs,Randall,UED}.  Almost all these models 
contain a heavy particle which shares the gauge quantum numbers of
the SM top quark, a `top partner', which leads to a 
relatively generic class of collider signals from their production and 
subsequent 
decay.
  One of the main motivations
 for introducing such a particle is to cancel the quadratically divergent 
contribution to the Higgs mass from the SM top, which has a large yukawa 
coupling to the Higgs.  This can come about using a scalar top partner, like
the stop $\tilde{t}$ in SUSY, or a fermionic one, like the $T'$ heavy 
top in Little Higgs models.  We would expect these particles to show up 
naturally at an energy scale of order $4\pi v$, where $v\approx 246$ GeV
is the Higgs field vacuum expectation value.\footnote{Notice that exceptions to this argument are certainly possible. For 
example, in Twin Higgs models \cite{twinhiggs} the particle that 
cancels the quadratic divergence of the SM top loop does not have 
the quantum numbers of the top.}.

A necessary requirement for a viable new
physics model is the suppression of the dimension-five and -six operators 
that are strongly constrained by the low energy data, such as
electroweak precision measurements, CP violation or flavor
changing neutral currents. In addition, dangerous
baryon/lepton number violating operators must be forbidden or strongly
suppressed. Motivated partly by these constraints and partly by being able
to provide a possible candidate for Cold Dark Matter, many new physics
scenarios incorporate a discrete symmetry under which the new physics
particles carry the opposite charge to SM particles. Typical
examples are R-parity in supersymmetry, KK-parity in UED, or $T$-parity in 
Little Higgs models \cite{LHT,tprime-model}.  
In such cases, assuming for minimality only a top partner (generically denoted
by $T$, unless otherwise specified) and a stable neutral particle which is 
the lightest 
parity-odd state ($A^0$), the predominant decay mode of the top partner is
\be
T \to t A^0 .
\ee
leading to the following hadron collider signal: 
\be
pp \to T\bar T\ X  \to t A^0\ \bar{t} A^0\ X \to t \bar{t} + \etmiss + X ,
\label{eq:signal}
\ee 
where $X$ represents the beam remnant and other possible hadronic activity,
and $\etmiss$ is the missing transverse energy. It would be desirable to be
able to distinguish between the different top partners that can give rise to 
this signal at the LHC.

The LHC will be a ``top factory":  About 80 million SM $t\bar{t}$ events will 
be produced from pure QCD, in addition to another 34 million single-top
events from the weak charged-current interaction for an integrated luminosity 
of 100 fb$^{-1}$. This provides a great opportunity
to study properties of top quarks in detail. However, SM top quarks 
will also serve as a non-trivial background for any
new physics signal with top quarks in the final state. Identifying
such signals above the huge background has been the focus of 
several recent  studies \cite{resonance-ttbar,tprime-study-1,tprime-study-2,
Nojiri}. 
In this article, motivated by the naturalness argument, we explore the signal of 
Eq.~(\ref{eq:signal}), concentrating on the semi-leptonic mode for the
 $t\bar t$ decays since the purely  hadronic top decay mode has been studied 
previously \cite{tprime-study-1,Nojiri}, with modest success.
We optimize the kinematical cuts
to separate the top partner signal from the SM backgrounds. 
In particular, we propose a reconstruction method for the top quark mass 
that allows it to take unphysical values to indicate the presence of physics
beyond the SM.
Furthermore, we comment on possible methods of determining the spin of the top
partner. Such measurements would be crucial in distinguishing
different underlying new 
physics scenarios, such as SUSY stop, or fermionic top partner pair production 
in UED or Little Higgs models with $T$-parity.  We survey possible
ways of getting  relevant spin information for the new particle, and
outline the difficulties involved.

The rest of the paper is organized as follows. In Sec.~II, we study the signal 
observability by carefully examining the kinematics and optimizing the
background suppression. In Sec.~III, we discuss the feasibility of spin
and mass determination. We summarize and conclude in Sec.~IV.

\section{Signal Observability}

In this section, we present a viable method for discovering new physics in
the $\tt + \etmiss$ final state as in Eq.~(\ref{eq:signal}).  
We assume for the purpose of the following discussion that the top 
partner $T$ is a 
color triplet under $SU(3)_C$ and a doublet under $SU(2)_L$. 

\subsection{Production Rates at the LHC}
\begin{figure}[tb]
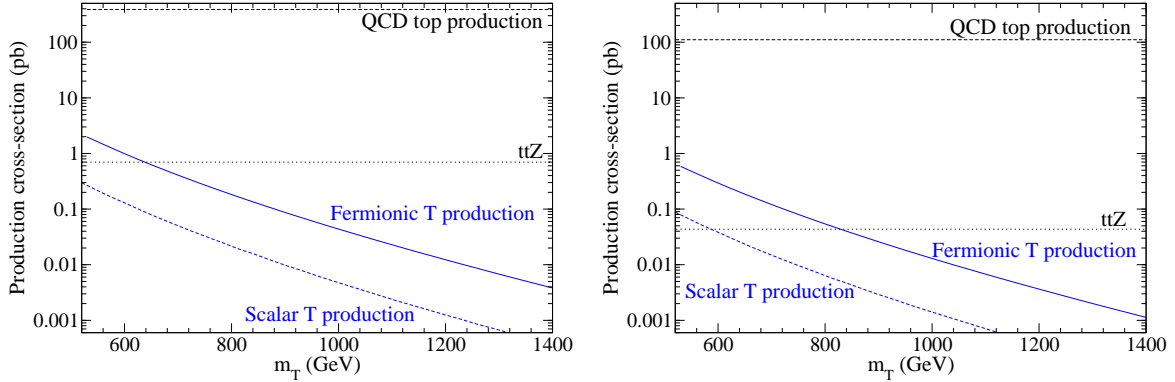

\begin{center}
\includegraphics[scale=0.29]{comparexsec.eps}
\hspace{0.05in}
\includegraphics[scale=0.29]{xsecbr.eps}
\caption{Leading order QCD cross section for top partner pair production
at the LHC, as a function of its mass. The solid line corresponds to a 
spin-$1\over 2$ particle,   
the dashed line to a spin-$0$ state. The two dashed horizontal
lines indicate the cross sections for the   
SM background processes $t\bar t$ and $t\bar t Z$ with tree-level
matrix elements.  The left panel shows the results 
before $T$ decay, and the right panel includes the decay branching
fractions to the  semi-leptonic final state $ bj_1j_2\ \bar b\ell^-
\bar\nu + \etmiss$, before any  
kinematical acceptance.}
\label{fig:sigmatot} 
\end{center}
\end{figure}

The leading production mechanism for the top partners is via QCD interactions
\be
q\bar q,\ gg  \to T \bar T.
\ee
In Fig.~\ref{fig:sigmatot}(a), we present the total leading order
$T\bar T$ production 
cross section at the LHC as a function of the mass of the $T$.  The solid line 
corresponds to a spin-$1\over 2$ particle; the dashed line corresponds 
to a spin-$0$ state. $\alpha_s$ is calculated at two loops, with the 
renormalization and factorization scales set equal to $\sqrt{s}/2$, and using
the CTEQ 4M parton distribution functions \cite{cteq4m}.  We see from the 
figure a factor of $8-10$ difference between the scalar and the fermion production
cross sections.  A factor of 4 comes from simple spin-state counting, and the 
remainder is due to threshold effects.

We will illustrate our procedure for background suppression using the example
of a fermionic top partner only. Since we expect that the phase space will be
the dominant factor in determining the kinematics, scalar top partner 
production and decay should have qualitatively the same behavior.

We assume that the fermionic top partner decays into a spin-1 neutral
stable particle  $A^0$ via a coupling: 
\be
\left(g_L \  \bar t \gamma^\mu P_L T + g_R \  \bar t \gamma^\mu P_R T\right)
A^0_\mu + h.c. 
\ee
Generically, we expect this coupling to be chiral ($g_L \neq g_R$). In
this study we choose $g_R=0$ although we do not expect that 
alternative choices will significantly change our optimization method.

We focus on the semileptonic channel, with one top decaying hadronically 
and the other decaying leptonically 
\begin{equation}
pp \to {\tt} \ A^0A^0 X \to bj_1j_2\ \bar b\ \ell^- \bar\nu \ A^0A^0\ X + c.c.
\label{totsigma}
\end{equation}
where the charged leptons are $\ell=e,\mu$. 
The signal thus consists of an isolated charged lepton, 
two $b$-quark jets and two light-quark jets plus large missing energy.
There are several advantages to studying this channel. First of all, the
branching fraction of the semileptonic mode is sizable, about
6 times larger than the cleaner purely  leptonic mode.
Secondly, although the hadronic mode has a branching fraction that is 1.5 times
larger still, 
the SM backgrounds for this mode are more severe than those for the 
semi-leptonic mode \cite{tprime-study-1, Nojiri}. 
Thirdly, one is able to distinguish the $t$ from the $\bar t$ using the charge
of the lepton in the final state. 
The cross sections for this channel including decay branching fractions as a
function of $m_T$  are shown in Fig.~\ref{fig:sigmatot}(b).

Based on event topology we expect several SM processes to constitute the
backgrounds to our signal. The leading SM background is QCD
production of top quark pairs:
\bea
pp \to \tt X \to  bj_1j_2\ \bar b\ \ell^- \bar\nu \ X + c.c.
\eea
The cross section for this process, depicted by the upper horizontal dashed lines in 
Fig.~\ref{fig:sigmatot}, is several orders of magnitude larger than our signal.

The next  background with a large missing energy is
\be
pp \to \tt Z X \to  bj_1j_2\ \bar b\ \ell^- \bar\nu \ \nu\bar\nu\ X + c.c.
\ee
with $Z\to \nu\bar\nu$. As seen in Fig.~\ref{fig:sigmatot} denoted by
the lower horizontal dashed lines,
even though the cross section for this process is smaller than the QCD
$t\bar t$ by about a factor of 600, its kinematics are more similar to that of
our signal, making such 
background events difficult to separate from the signal.  
We will elaborate on this below.

Another large SM background is from a process with no top quark
\be
pp \to W b\bar b\  j_1j_2\to \ell^- \bar\nu\  b \bar b\ j_1 j_2\ X + c.c.
\ee
In our study, we simulate the SM $t\bar{t}$ and $t \bar{t} Z$ backgrounds
using PYTHIA \cite{pythia}, while we use ALPGEN \cite{alpgen} for 
the $W(\rightarrow \ell \nu)\ bb\ jj$ background. 
We perform all calculations at parton level. With the stringent 
acceptance cuts to be discussed below, we expect that all next-to-leading 
order QCD effects, such as hadronization and initial and 
final state radiation, will not alter our results appreciably.

\subsection{Extracting Top Partner Signal}

\begin{figure}[tb]
\begin{center}
\includegraphics[scale=0.4,angle=270]{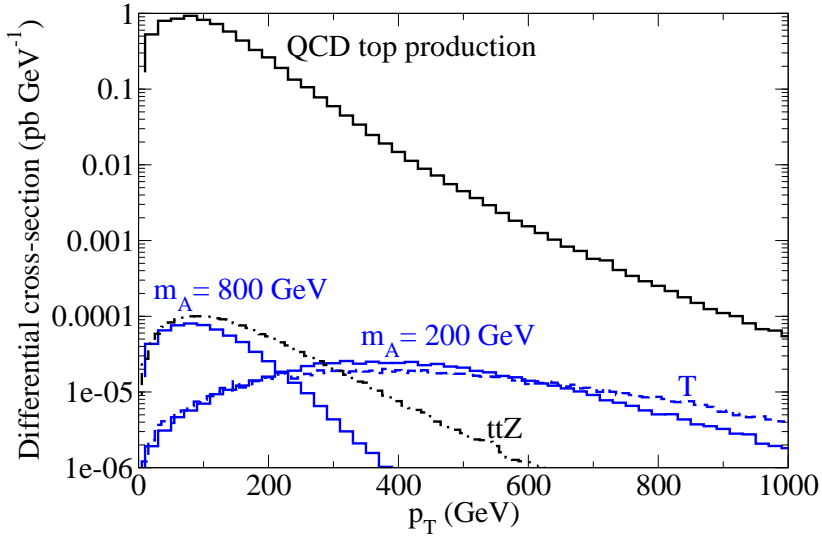}
\caption{Transverse momentum distributions for the top quark from QCD
$\tt$ production (the top curve), $\tt Z$ production  (dot-dashed), 
and from $T$ decays for  
$m_T=1$ TeV and $m_A=800,\ 200$ GeV, respectively. For comparison,
we include the $p_T$ of the fermionic $T$ (long-dashed).}
\label{fig:pt} 
\end{center}
\end{figure}

In this section we present our main results on separating signal events 
from backgrounds.  
Although the SM backgrounds to the heavy top partner pair production and 
semileptonic decay are substantial,
there are many kinematical differences between them that can be
exploited. Moreover, we can take full advantage of the observation that 
the neutrino from $W$ decay is largely responsible for the $\etmiss$ in 
the $t\bar{t}$ and $Wbbjj$ backgrounds, while it contributes only a fraction 
of $\etmiss$ in the signal. 

A crucial parameter for the signal kinematics is the mass difference
between $T$ and $A$, 
\be
\dm \equiv m_T - m_A.
\ee
The energy of the top quark from $T$ decay is 
$E_t\approx 0.5(1+m_A/m_T) \dm $ in the rest frame of the $T$. 
For a sufficiently large mass difference, the top quark can be very energetic.
For a small $\dm$, however, the top quark has little kinetic energy and the 
signal  kinematics are very similar to 
those of the $\tt$ background. We will present results for two benchmark 
scenarios for illustration
\be
m_T=1\ {\tev},\quad {\rm and}\quad m_A=200,\ 800\ {\gev}.
\ee
In Fig.~\ref{fig:pt}, we show the transverse momentum distributions  for 
the heavy $T$ (dashed line), for the top quark from $T$ decay for our two benchmark values,
and from QCD $t\bar t$ (solid line), and $t\bar t Z$ (dot-dashed line).  
>From this graph one can see that the $p_T$ spectrum of the heavy $T$ 
has the expected broad plateau near $(0.3-0.6) m_T$. 
The $p_T$ spectrum of the top quark from $T$ decay for small $m_A$ is 
similar to that of the $T$ quark itself, while for the small mass difference
case ($m_A=800$ GeV), it is more similar to the $t\bar t$ background.

To simulate  the detector acceptance 
\cite{unknown:1999fr,unknown:2006cm}, we first impose the basic cuts
\bea
&& p_T^\ell > 20\ {\gev},\quad |\eta_\ell | < 2.5,\quad \Delta R_\ell > 0.3,\\
&& E_T^j > 25\ {\gev}, \quad |\eta_j | < 2.5,\quad \etmiss > 25\ {\gev},\\ 
&& E_T^b > 30\ {\gev}, \quad |\eta_b | < 2.5,\quad \Delta R_j,\
\Delta R_b >0.4, 
\eea
We adopt  relatively small isolation cuts 
in order to accommodate the kinematics of a fast-moving top quark from 
a heavy $T$ decay.
We simulate the calorimetry responses for the energy measurements
by adopting Gaussian smearing \cite{unknown:1999fr} with the following
parameters:
\bea
{\Delta E_e\over E_e} =  {10\%\over \sqrt {E_e (\gev)}} \oplus 0.7\%,\qquad
{\Delta E_j\over E_j} =  {50\%\over \sqrt {E_j (\gev)}} \oplus 3\%.
\eea
We do not separately smear the muon momentum. In the energy-momentum
range of current interest, the lepton resolutions should not make 
any appreciable difference in the results since the dominant effect is from
hadronic smearing. 
We require two tagged $b$-jets in the selection. In our presentation
of the discovery reach, we use a $b$-tagging efficiency of
\cite{unknown:1999fr,unknown:2006cm} 
\be
\epsilon_b = 60\% ,
\ee
which is appropriate for the range of $p_T^b$ that we are interested in at
low luminosity. 
\begin{figure}[tb]
\includegraphics[scale=0.29,angle=270]{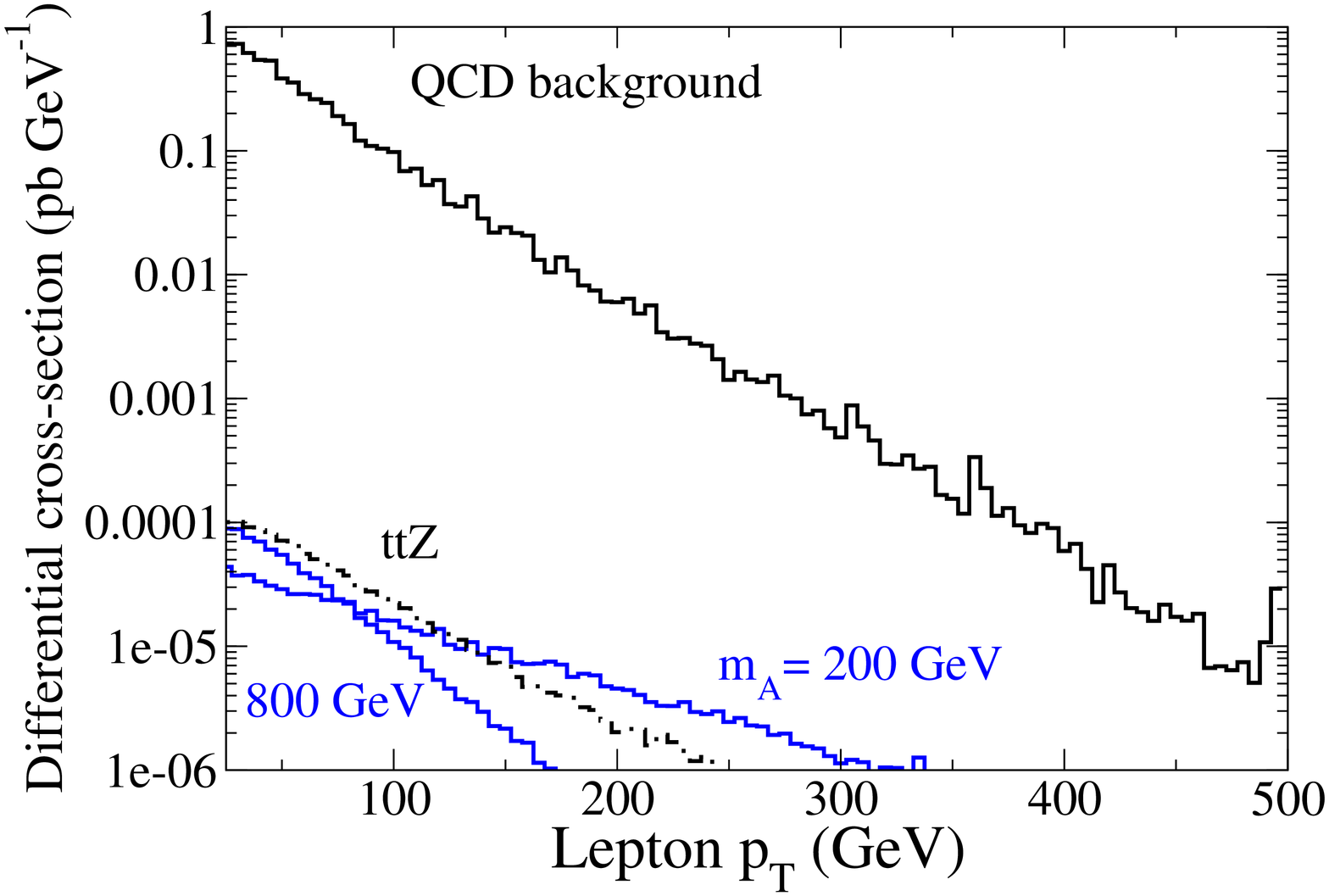} 
\includegraphics[scale=0.29,angle=270]{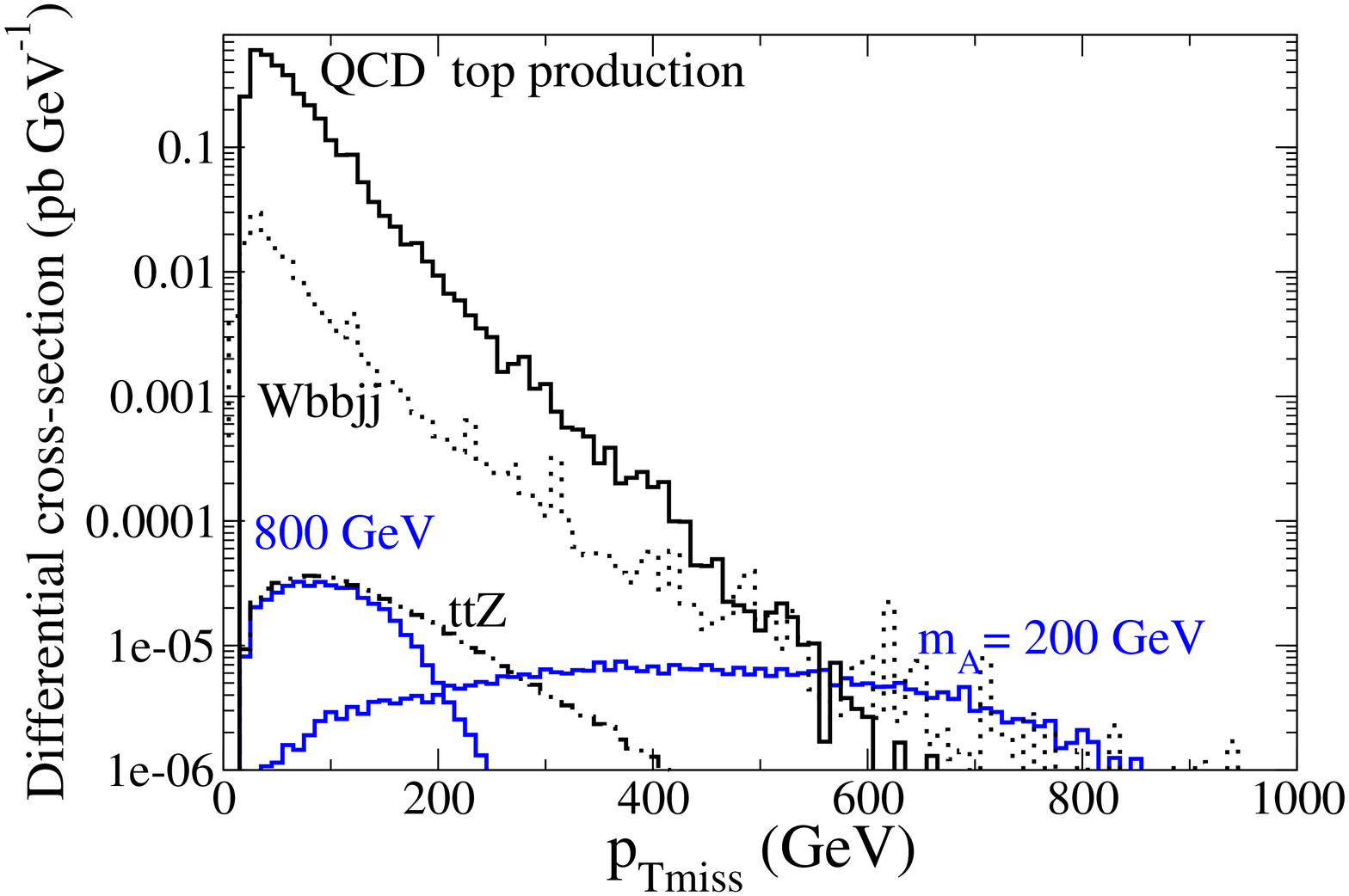} 
\begin{center}
\includegraphics[scale=0.29]{meff.eps}
\caption{Differential distributions for (a) the transverse momentum of 
the charged lepton $p_T(\ell)$, (b) 
 the missing transverse momentum $p_T^{miss}=\etmiss $, and 
 (c) the effective transverse mass of the final state system, respectively.}
\label{fig:TM}
\end{center}
\end{figure}
In Fig.~\ref{fig:TM} we present some characteristic kinematical distributions 
for the signal and backgrounds with basic cuts imposed.  
The transverse mass variable $M_T^{\rm eff}$ is defined in 
Appendix~\ref{sec:def_meff}.
The heavy $T$ signal generically leads to energetic decay products unless 
the mass difference $\dm$  becomes very small.
>From Fig.~\ref{fig:TM}(b),  we see that a large missing energy cut of 
$\not{\! \! E}_T > 350$ GeV  could be imposed to effectively remove the $\tt Z$ 
backgrounds, but such a requirement 
will eliminate the signal in the case of small mass splitting $\Delta
M_{TA} \sim 200 - 300$, where 
$A^0$ only carries away a small amount of kinetic energy
$\sim 0.5 (1-m_A/m_T)(\dm - m_t)$. 
As seen in Fig.~\ref{fig:TM}(c), the effective transverse mass does not 
provide more discriminating power than $\etmiss$. There are in
principle other transverse variables one could use to distinguish signal
from background, such as the cluster transverse mass \cite{Barger:1983jx}
 (defined in Appendix~\ref{sec:def_meff}), or $M_{T_2}$ \cite{mt2}. However, 
these variables are largely similar, and unlikely to do significantly better
than the $\etmiss$ and $M_T^{\rm eff}$ variables presented here. 

\begin{figure}[tb]
\includegraphics[scale=1.0]{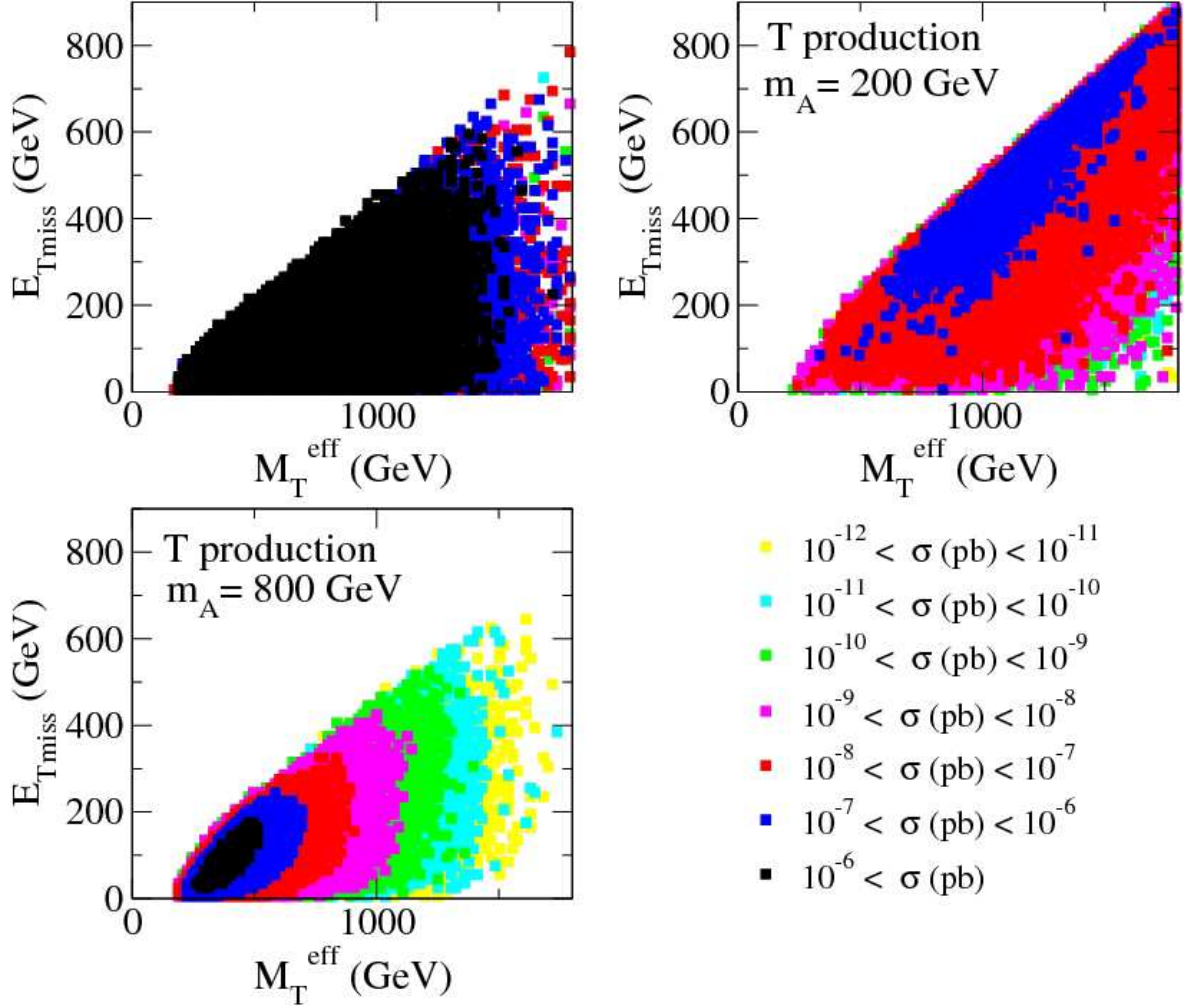}
\caption{Kinematical correlation of events between $\etmiss$ and
  $M^{\rm eff}_T$ for  
(a) the QCD $t\bar t$ background,
(b) and (c) the $T$ signal with $m_A=200,\ 800$ GeV respectively. 
The color codes indicate the size of the cross sections.}
\label{fig:etmt}
\end{figure}

As recently suggested in \cite{Nojiri}, one may consider exploring the 
correlation between $\etmiss$ and $M^{\rm eff}_T$, which we present in 
Fig.~\ref{fig:etmt} for  (a) the QCD $t\bar t$ background,
(b) and (c) $T\bar T$ pair production with $\Delta M_{TA}=200,\ 800$
GeV respectively. Two remarks are in order. First, the correlation is
more distinctive between the signal and background when the mass difference 
$\dm$ is large as seen in Fig.~\ref{fig:etmt}(b), namely 
$M^{\rm eff}_T \sim 2 \etmiss \sim 2 m_T$. 
It tends to be very similar to the $\tt$ background distribution when 
$\dm \sim m_t$ as in  Fig.~\ref{fig:etmt}(c). This less desirable
situation was not considered 
in  \cite{Nojiri} due to their parameter choice in favor of a dark
matter candidate, in the context of a particular model \cite{LHT}.
Second, due to the overwhelmingly large  rate
of the $\tt$ background, this correlation variable alone is not
sufficient to separate  
the signal in the semi-leptonic channel, as seen for the integrated rates
by the color codes in Fig.~\ref{fig:etmt}. 

\begin{figure}[tb]
\begin{center}
\includegraphics[scale=0.4,angle=270]{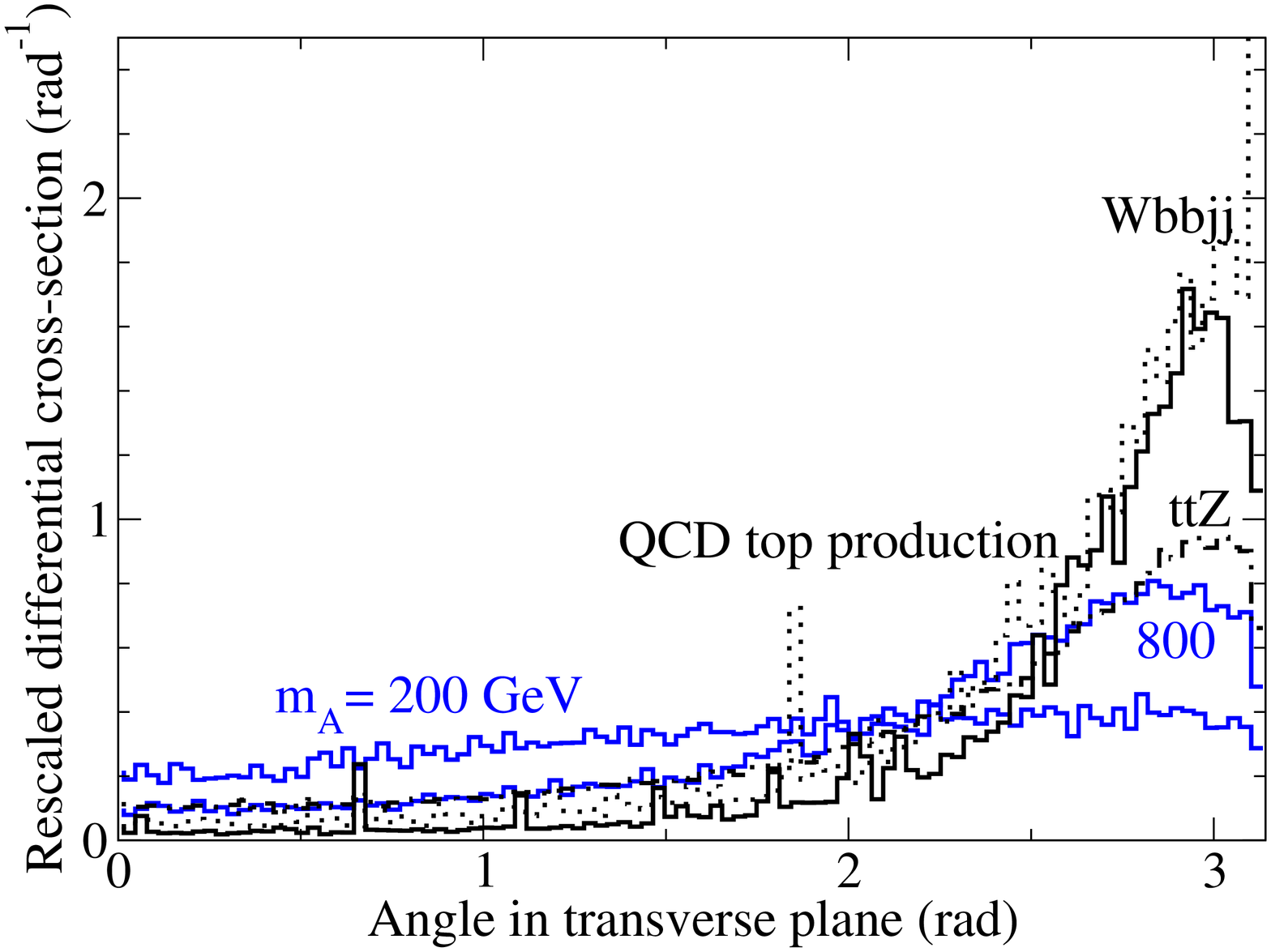}
\caption{Normalized opening angle distributions in the transverse
plane $\phi_{t-bl}$ for the signal $m_A=200,\ 800$ GeV (solid), 
and the $t\bar t$ background (solid), $Wb\bar b jj$ (dashed),
and $t\bar t Z$ (dot-dashed). }
\label{fig:angle}
\end{center}
\end{figure}
There are other kinematical features that one could utilize to separate the
signal from the backgrounds.
One such variable is the transverse angle between the $t$ and $\bar t$: 
We expect $t$ pairs from pure QCD production to be co-planar in the $2\to 2$ 
scattering plane and thus back-to-back in the transverse plane, while those 
originating from $T\bar T$ 
decays will be more randomly oriented due to the transverse kicks of the missing
$A^0$ particles.  
Although it is difficult to observe this co-planarity by studying the decay
products of the top quarks due to the presence of the neutrino, 
one could still exploit the back-to-back nature of the $t\bar t$ background 
by evaluating the opening angle in the transverse plane between the hadronic 
top and the reconstructed momentum $\vec p_{bl} = \vec p_b+\vec p_l$,
called $\phi_{t-bl}$. For the $Wb \bar{b} jj$ background, we
  followed the same reconstruction procedure as the signal and other
  backgrounds (more details below) with $\phi_{t-bl}$ in this case 
defined by the angle between hadronic ``top'' and the unused $\vec p_{bl}$. 
We plot this distribution, rescaled by the total cross section, in Fig.~\ref{fig:angle}.  
Indeed we see a strong correlation for $\tt$, $W b\bar b jj$ and $t\bar tZ$, 
while the distribution for the signal is rather flat, especially
when $\dm$ is sizable. 
The discriminatory power of this variable decreases as the mass splitting 
becomes smaller.  In addition this correlation will become less pronounced on 
inclusion of realistic QCD radiation and parton showering.
We will therefore not devise 
a specific cut on this variable, although this could be implemented
in a full optimization procedure.

Another obvious variable we might cut on is the di-jet mass, which
will reconstruct a $W$ for the signal, but not so for the background
$W b\bar b jj \to \ell\nu\ b\bar b jj$. This has been used in the top quark
signal analysis at the Tevatron and we will take full advantage of this 
fact in our analysis.

\begin{figure}[tb]
\begin{center}
\includegraphics[scale=0.45,angle=270]{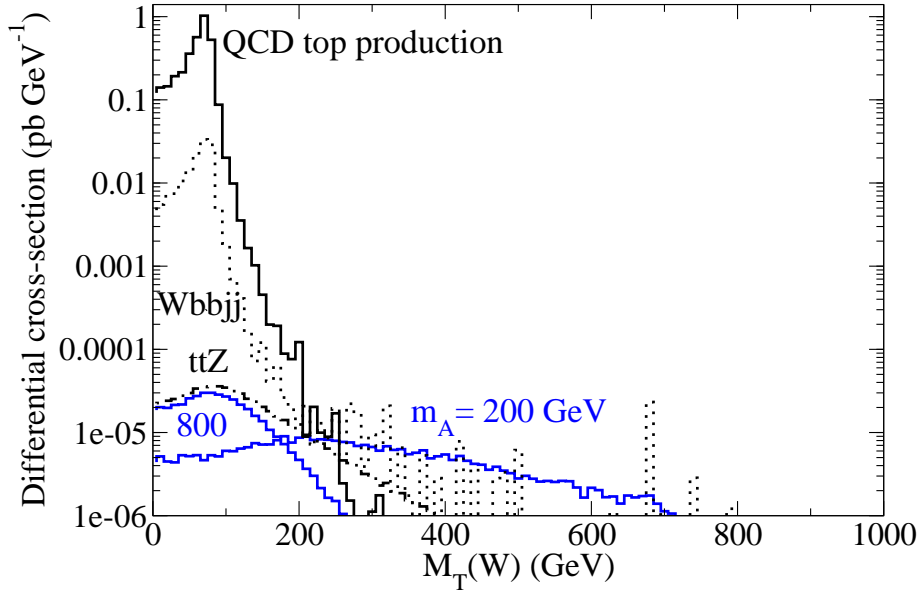}
\caption{$M_T(W)$ distribution for both signal and background. The
  peak near $M_W$ is prominent for the $t \bar{t}$ and $W(\rightarrow \ell
  \nu )bbjj$ backgrounds. }
\label{fig:mwt} 
\end{center}
\end{figure}

Next, we turn to a discussion of reconstructing intermediate particle masses
using kinematical variables.
The first variable one might reconstruct is the transverse mass of
the leptonically-decaying $W$ by assuming $p_{\nu T} = E_{\nu T}=\ptmiss$. 
As usual 
we define the transverse mass for the leptonic system as
\be
M^2_T(W)= (E_{\ell T}+ E_{\nu T})^2- (\vec{p}_{\ell T} + \vec{p}_{\nu T})^2~~. 
\ee
When the missing energy comes only from the single 
neutrino from $W$ decay,  we expect to see a Jacobian peak 
$M_T(W) \sim M_W$, but this will not be so for the signal events where
the missing energy is carried away by two massive particles in addition to
the neutrino. 
This reconstructed variable is plotted in Fig.~\ref{fig:mwt} for the signal 
and backgrounds. 
We see  that a transverse mass cut, for
example $M_T(W)> 220$ GeV, on the
leptonic products can effectively suppress the QCD $t\bar t$ and $Wbb jj$
backgrounds  
due to the presence of $W$ production and its subsequent leptonic decay.

Note that we have not thus far taken advantage of the fact
that the kinematics of SM $\tt $ in the semi-leptonic channel is fully
reconstructible \cite{top-recon-exp}, following a well-known procedure in 
which the on-shell condition for the $W$ boson is used to determine the 
four-momentum
of the massless neutrino, upto a two-fold ambiguity. Imposing the on-shell
condition for the leptonically-decaying top eliminates this ambiguity, while
the mis-pairing for $b$ and $\bar b$ can be
simultaneously reduced by minimizing the sum of the differences between the
masses of the reconstructed tops and the actual top mass 
\be
\left(m_t-m(b_1jj)\right)^2 + \left(m_t-m(b_2 \ell\nu) \right)^2.
\ee
Our detailed reconstruction procedure is summarized in
Appendix~\ref{top-recon}.

The situation is expected to be very
different for the new physics signal in which the $\tt $ system
recoils against the missing massive particles.  
Applying the same reconstruction technique to the signal events, one
would therefore not expect to reconstruct the top quark successfully since the
$\etmiss$ contains a large contribution from the momenta of the $A^0$s, which
cannot be measured independently. This
typically results in either not being able to find 
physical solutions for the longitudinal component of the
missing energy, or in a reconstructed mass that is very different from $m_t$.
Therefore, this naive reconstruction procedure provides us with 
an effective way to distinguish the signal from the $t\bar t$ background. 
We encode the effect of unphysical momentum solutions by 
allowing the reconstructed top quark mass $m_t^r$ to carry an imaginary 
part (see Appendix~\ref{top-recon} for details). In this case, a large unphysical, {\it i.e.} 
imaginary, value for the reconstructed mass is a signature of events with 
missing particles beyond a single massless neutrino.  

\begin{figure}[tb]
\begin{center}
\includegraphics[scale=0.6]{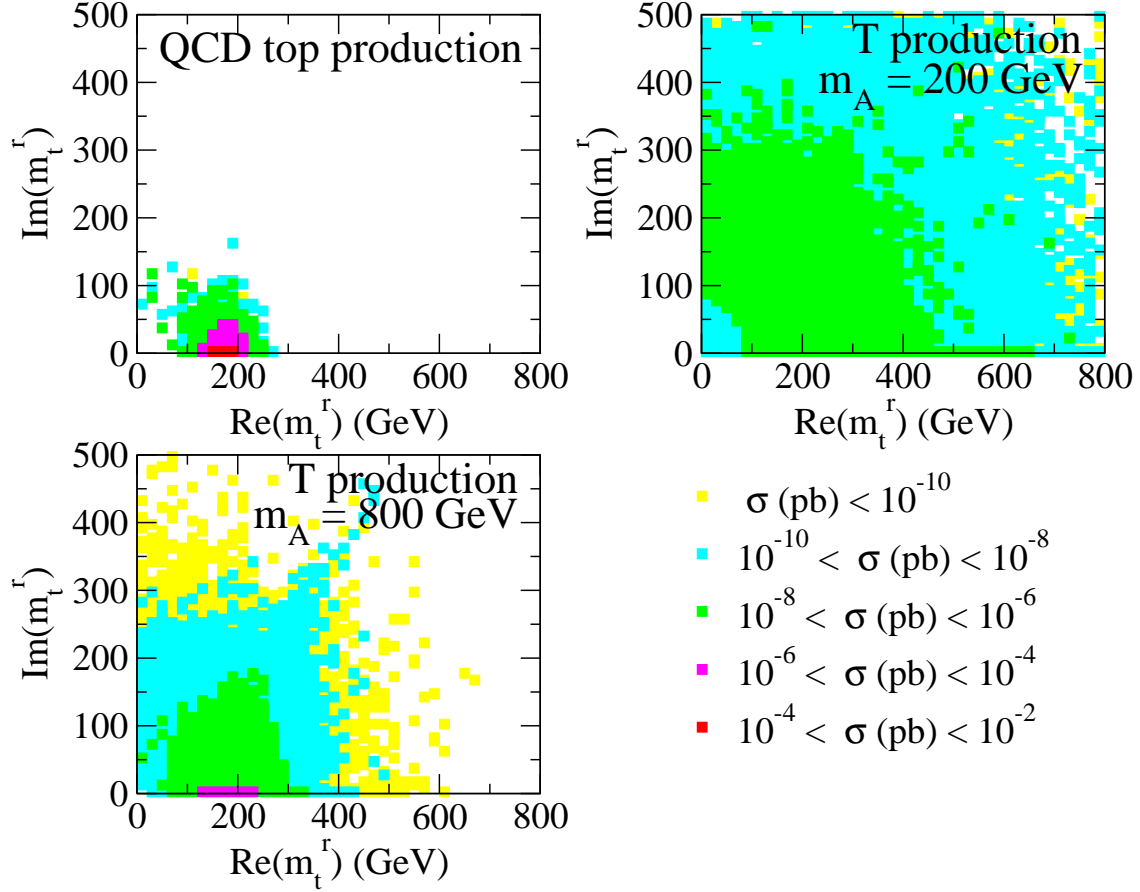}
\caption{The reconstructed mass of the leptonically decaying
top quark in the complex plane (a) for $t\bar t$
background, (b) for $m_A=200$ GeV and (c) for $m_A=800$ GeV. Allowing this
variable to take on complex values serves as a pointer to new physics, an
imaginary value is a signature of events with new missing particles.}
\label{fig:tmass} 
\end{center}
\end{figure}

The results for the reconstructed $m_t^r=m(b_2\ell\etmiss)$ are shown in the complex
plane in Fig.~\ref{fig:tmass}. As seen in 
Fig.~\ref{fig:tmass}(a), the reconstructed mass for the $t\bar t$ background 
is highly concentrated near $m_t$ on the real axis 
although there are still a small number of events that give an unphysical top
quark mass 
due to the energy-momentum smearing effects of the detectors. For the
signal events, it is spread out over a large region as seen in Fig.~\ref{fig:tmass}(b) and (c).
We are thus motivated to impose a cut on $|m_t - m_t^r|$.  The choice
\be
|m_t - m_t^r| > 110\ {\rm GeV} 
\label{eq:mtcut}
\ee
for example, essentially eliminates the $t\bar t$ background.
The range of unphysical values is reduced for the signal
when the mass difference $\Delta M_{TA}$ becomes smaller, confirming the
fact that when the missing
$A^0$s carry little kinetic energy, there is effectively no difference 
between the kinematics of the  
signal and $t\bar t$ background. 
 In this kinematical regime the $A^0$ moves slowly and its
momentum does not contribute significantly to $\etmiss$. Therefore
we should be able to approximately reconstruct $m_t$ by assuming
$p_{\nu T} = \ptmiss$. 

\begin{table}[t]
{\footnotesize
\begin{tabular}{c |c c c | c c c || c c c}
\hline\hline
 & \multicolumn{3}{c}{$~S:~m_A=$200 GeV~~} & \multicolumn{3}{c}{$m_A=$800
 GeV~} & $B:~\tt~~$ & $\tt Z~~$ & $Wbbjj$ \\
\hline 
& eff.  & $S/B$ & $S/\sqrt{B}$ & eff. & $S/B$ & $S/\sqrt{B}$ &
 eff. & eff.  & eff. \\
\hline
Basic cuts & 0.28 & $10^{-4}$  & 0.2 & 0.32 & $10^{-4}$ & 0.2 & 0.24
& 0.29 &  $-$ \\ 
\hline
$\etmiss>350$& 0.65 & 0.1 & 4.3 & $5\cdot 10^{-4}$ & $8\cdot 10^{-5}$ & $4
\cdot 10^{-3}$ & $6 \cdot 10^{-4}$ & 0.03 &  $7\cdot 10^{-3}$ \\
$\etmiss>600$& 0.22 & 1.0 & 8.6 & $9 \cdot 10^{-8}$ & $4 \cdot
10^{-7}$ & $4 \cdot 10^{-6}$ & $2 \cdot 10^{-6}$ & $2 \cdot 10^{-3}$ &  $8 \cdot 10^{-4}$ \\
$|m_{jj} - M_W| < 20$ & 0.97 & $10^{-4}$ &0.2 &0.95 & $10^{-4}$& 0.2  & 0.96&0.89 & 0.11 \\ 
$120 < m_t^{had} < 180$& 0.76& $10^{-4}$& 0.2 & 0.73&$10^{-4}$ &0.2 &0.77 &0.72 & 0.10 \\ 
$\phi_{t-b\ell}<2.5$ & 0.75 & $2\cdot 10^{-4}$ & 0.3 & 0.54 & $2\cdot 10^{-4}$ & 0.2 & 0.26 & 0.50 &  0.31 \\
$M_T (W) >220 $ &0.62 & 0.7& 13&0.03 &$4\cdot 10^{-2}$ &0.7 &$2\cdot 10^{-5}$ &0.11 & $2\cdot 10^{-3}$\\ 
$|m_t^r -175|>110$ & 0.75 & $8\cdot 10^{-3}$ & 1.5 & 0.08 & $1\cdot 10^{-3}$ & 0.2 & $5\cdot 10^{-5}$ & 0.17 &   0.30\\
\hline\hline
\end{tabular}
}
\caption[]{Effect of individual kinematical cuts on the signal for $m_T=1$ TeV 
and backgrounds. All non-detector efficiencies are calculated for events which pass
the basic cuts; masses and energies are in GeV.  The statistical
significance ($S/\sqrt B$) is computed for a luminosity of 100 $\fbi$. }
\label{tab:sb}
\end{table}
To summarize the discussion in this section, we present in
Table~\ref{tab:sb} the signal and background efficiencies after applying 
individual cuts, for all events that have passed the basic cuts.  The table
also includes the signal-to-background ratios $S/B$ and statistical 
significance for signal observability $S/\sqrt{B}$ for the individual cuts for
an integrated luminosity of 100 fb$^{-1}$.  From this we can see, for example,
that for  $m_T = 1$ TeV, which has a production rate of about $0.045$ pb at the
LHC, a combination of an $\not{\! \! E}_T$ cut and reconstructions of $m_{jj},\ M_T(W)$
and the top mass can significantly reduce the background.

\subsection{Discovery Reach}

In this section we present the observational reach for our signal 
$T \bar{T} \rightarrow t \bar{t} A^0 A^0 \rightarrow \ell bbjj + \etmiss$. 
We will choose our cuts based
on the kinematical and reconstruction variables studied in the previous
section. The particular choices we make here are designed to
illustrate the potential of enhancing the signal-to-background ratio. A
complete optimization of kinematical cuts can be based on
our variable studies, but is not performed in this work.

Although the reconstruction of $m_t^r$ is very effective in suppressing
the $t\bar t$ background and distinguishing the signal, it does not provide
a discrimination against the $Wbbjj$ background which does not have
a real top quark to begin with. We thus impose additional cuts to reduce
this large background
\bea
&& 70 \ {\gev} < m_{jj} < 90 \ {\gev},\\
&& 120 \ {\gev}  < m_{t}^r|_{\rm had} = m(b_1jj)< 180 \ {\gev},
\label{eq:Wcuts}
\eea
where two $b$-tags are required for reconstruction. 

The choice of an $\etmiss$ cut is more involved. 
As discussed previously , on the one hand
imposing an appropriate $\etmiss$  cut will
definitely help suppressing the background ($\tt Z$ in particular).
On the other
hand a large $\etmiss$ cut will eliminate the signal for a small mass splitting.  
Hence we optimize the search by making a variable $\etmiss$ cut 
\be
\etmiss > 0.2 \times \dm. 
\ee
We also impose the reconstructed leptonic top mass cut detailed 
in Eq.~(\ref{eq:mtcut}). 

\begin{figure}
\begin{center}
\includegraphics[scale=0.45]{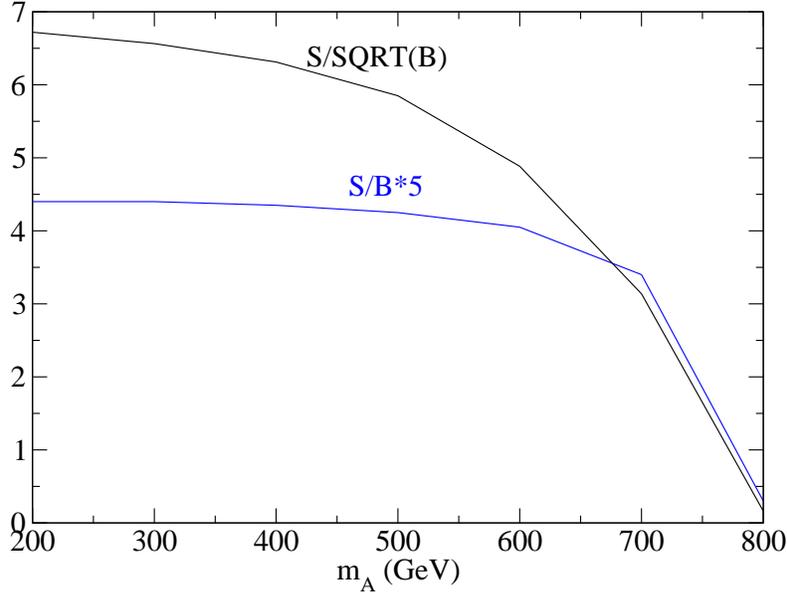}
\caption{Signal-to-background ratio and the 
statistical significance (100 fb$^{-1}$ integrated luminosity)
as a function of $m_A$ for a fermionic top partner of mass $m_T=1$ TeV.}
\label{fig:reach}
\end{center}
\end{figure}

In Fig.~\ref{fig:reach}, we present the signal-to-background 
ratio and statistical significance as a function of 
$m_A$ for a fermionic top partner of mass $m_T=1$ TeV and 
with 100 fb$^{-1}$ integrated luminosity.
We see that after imposing our proposed combination of cuts 
we have significant signal observability for a mass
of upto $m_A\approx 750$ GeV, which corresponds to $\Delta M_{TA}\sim 250$ GeV. 
For an even smaller mass difference $\Delta M_{TA} \sim m_t$, 
it is still challenging to select out the signal.

\psfrag{MT}{m$_{\rm{T}}$}
\psfrag{MAH}{m$_{\rm{A}}$}

\begin{figure}
\begin{center}
\begin{tabular}{cc}
\includegraphics[scale=0.6]{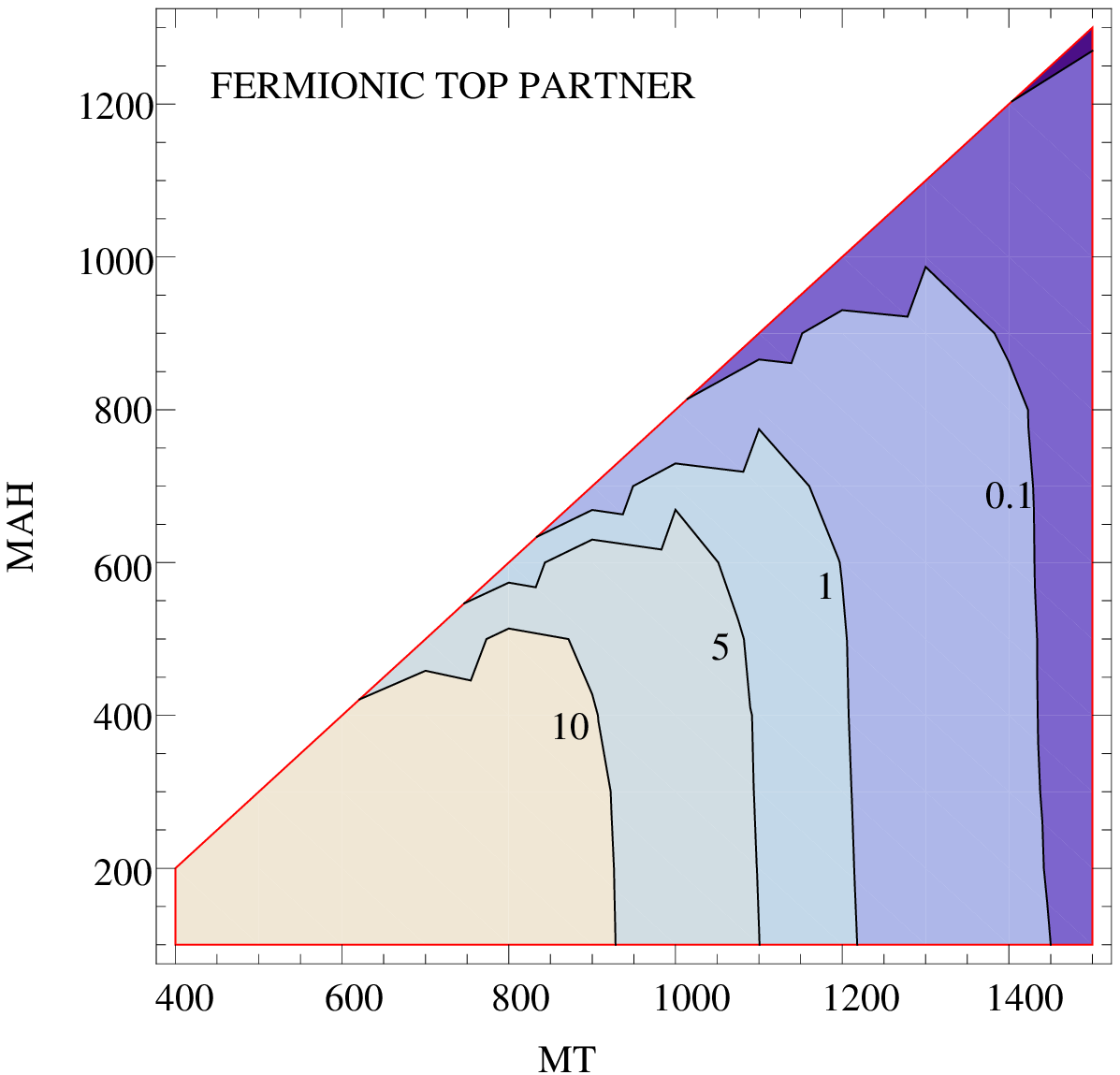} &
\includegraphics[scale=0.6]{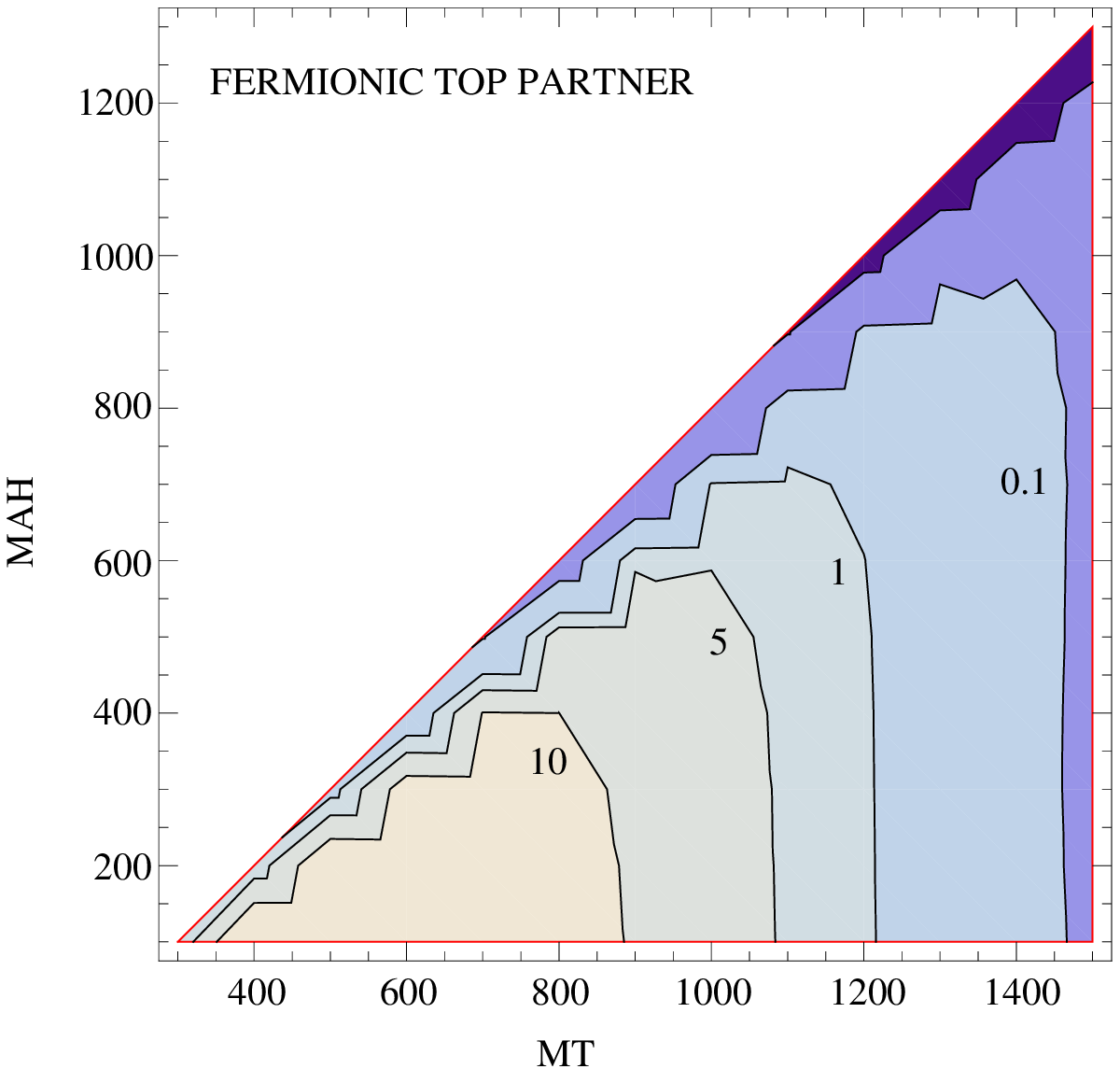} 
\end{tabular}
\end{center} 
\caption{Contours of  statistical significance for a fermionic top partner with 100 $\fbi$ of luminosity 
in the $m_T-m_A$ plane, the left-handed panel with the cuts described in this section;
the right-hand panel  with an additional cut on transverse mass $M_T (W) > 220$ GeV. }
\label{fig:MTDM}
\end{figure}

We now study the more comprehensive reach in the two-parameter $m_T$-$m_A$ plane 
with combined cuts. Our results are presented in 
Fig.~\ref{fig:MTDM} in the form of contour plots of signal observability
for an integrated luminosity of 100 fb$^{-1}$, with $10\sigma,\ 5\sigma,$ and $\ 1\sigma$ 
contours shown.
In the left-hand panel we implement the cuts described in this section, 
while in the right-hand panel we include an additional cut on the $W$
transverse mass $M_T (W) > 220$ GeV. This
additional cut does not enhance 
the signal significance ($S/\sqrt{B}$) appreciably due to correlations with 
other cuts, in particular $m_t^r$.  It does however improve the signal
to background ratio ($S/B$), and hence helps control systematic effects.
We also note that the reach here is similar to that
obtained in the fully hadronic mode \cite{tprime-study-1}. 
\psfrag{MT}{m$_{\rm{T}}$}
\psfrag{MAH}{m$_{\rm{A}}$}
\begin{figure}
\begin{center}
\begin{tabular}{cc}
\includegraphics[scale=0.6]{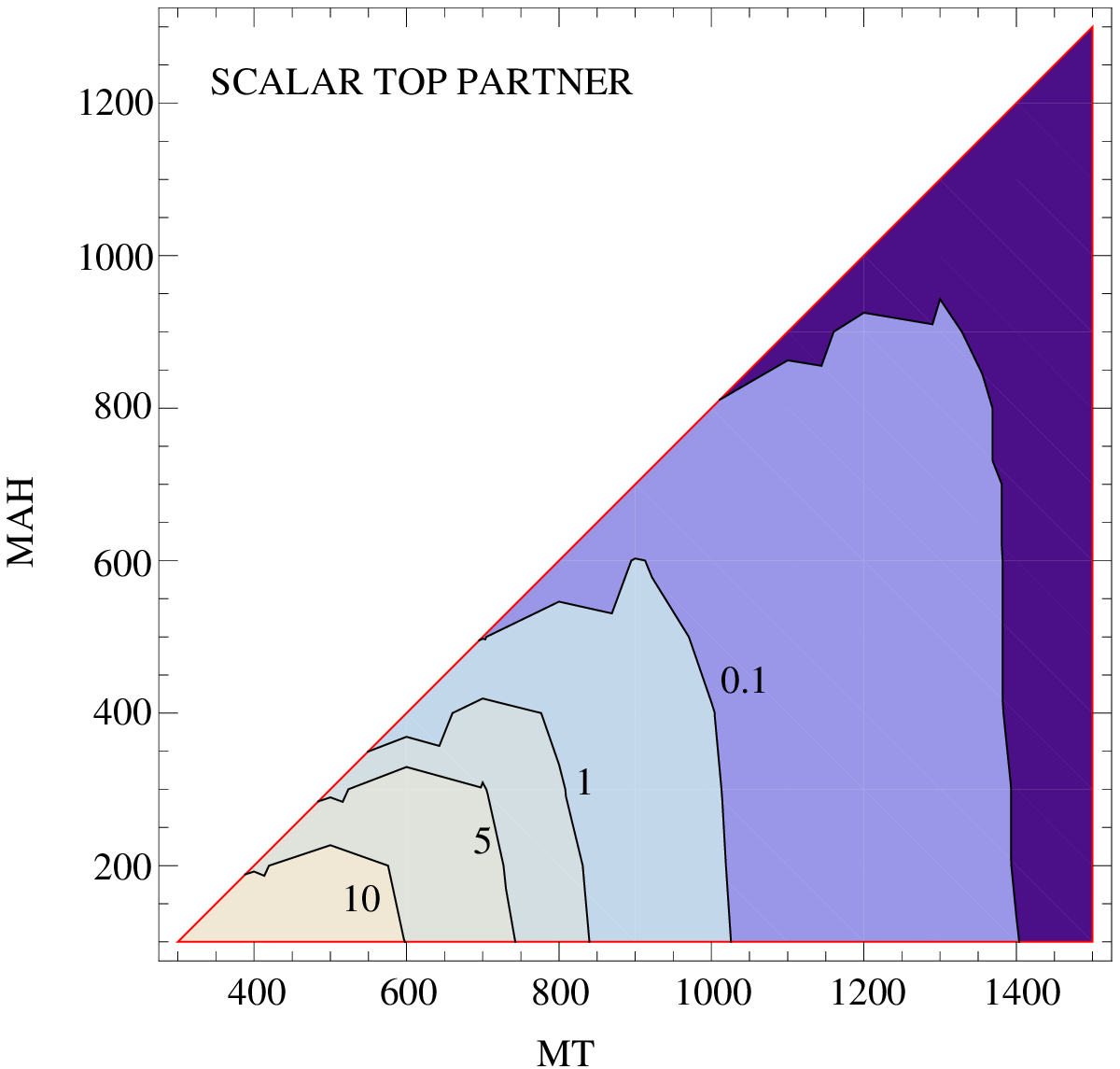} &
\includegraphics[scale=0.6]{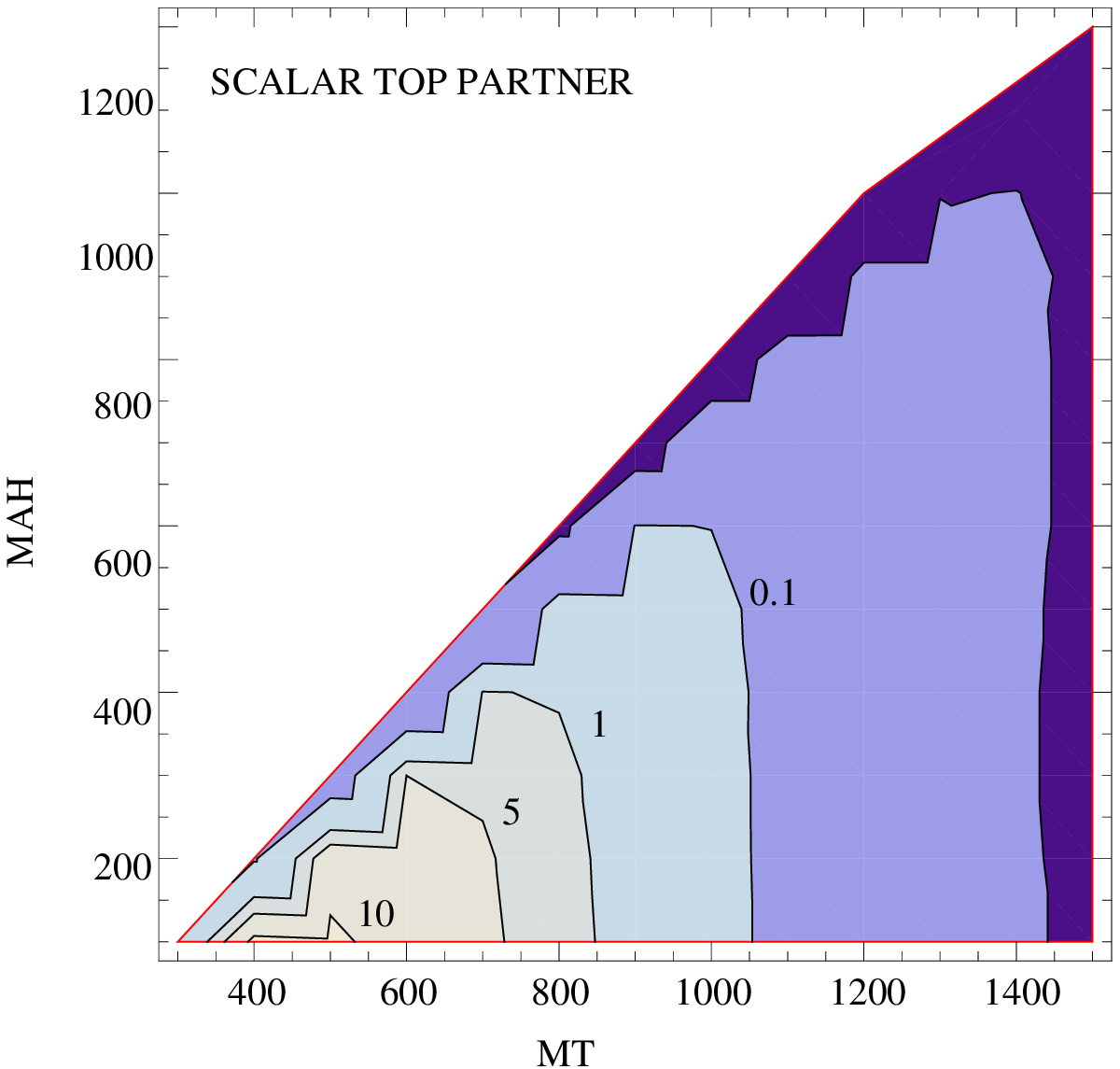} 
\end{tabular}
\caption{Contours of  statistical significance for a scalar top partner,
with 100 $\fbi$ of luminosity
in the $m_T-m_A$ plane. The panel on the right 
is the result with an additional cut on transverse mass $M_T(W) > 220$ GeV. }
\label{fig:MSTDM}
\end{center} 
\end{figure}
It is straightforward to extend our results to the case of the scalar top
partner decaying to $t A^0$ (e.g. $\tilde{t}_R\to t  \chi_0$ in SUSY).  
As in 
the case of the fermionic top partner, 
the crucial parameter controlling much of the kinematics is
the mass splitting $\dm$=$m_{\tilde t}^{} - m_{\chi^0}^{}$. Given the
same mass splitting, we expect the kinematics will be quite 
similar to the case of fermionic $T$.~\footnote{There are some subtle differences
  \cite{tprime-study-1}, which we will comment on in
  Sec.~\ref{sec:spin}. We do not expect such differences to 
  affect the discovery reach significantly.} 
  Therefore, we should
expect any difference in reach to be mostly due to the lower production
cross section for the scalar top partner. 
The reach, with and without a transverse
mass cut, is shown in Fig.~\ref{fig:MSTDM}.

\section{Remarks on Distinguishing a Scalar Top Partner from a Fermion }
\label{sec:spin}

As argued above, the hadron collider signatures of a fermionic top partner are
expected to be very similar to those of scalar, making it 
challenging to directly measure the spin of
the top partner at the LHC. We discuss several approaches to tackling this 
problem in this section.

The most straightforward way to tell the fermionic partner from the scalar
would be by using the difference in their production rates, as seen in
Fig.~\ref{fig:sigmatot}. 
However, one cannot interpret this without making some assumption about the 
underlying model, since several degenerate
scalars could be produced with the same total rate as a Dirac fermion of the same
mass, for example. Even given some set of reasonable assumptions, 
as pointed out in Ref.~\cite{tprime-model, tprime-study-1}, 
we are left with the problem that masses cannot be measured directly 
(at least in the situation where there is more than one type of particle
contributing to the missing energy) since most known observables only measure 
the mass difference.
Consequently, a cross section
measurement alone will not determine the particle's mass or spin.
Recently, several new methods have been proposed to tackle the
problem of measuring absolute mass scale
\cite{mass-methods-1,mass-methods-2}. However
the reconstruction method in Ref.~\cite{mass-methods-1} relies on longer 
decay chains with more kinematical handles, 
and therefore is not directly applicable in our
case.  Application of the method in Ref.~\cite{mass-methods-2}, which
relies on accurately measuring the end point of the $m_{T_2}$ variable,
to the case of minimal top partner decay has yet to be studied in this context. 

\begin{figure}
\begin{center}
\includegraphics[scale=0.45,angle=270]{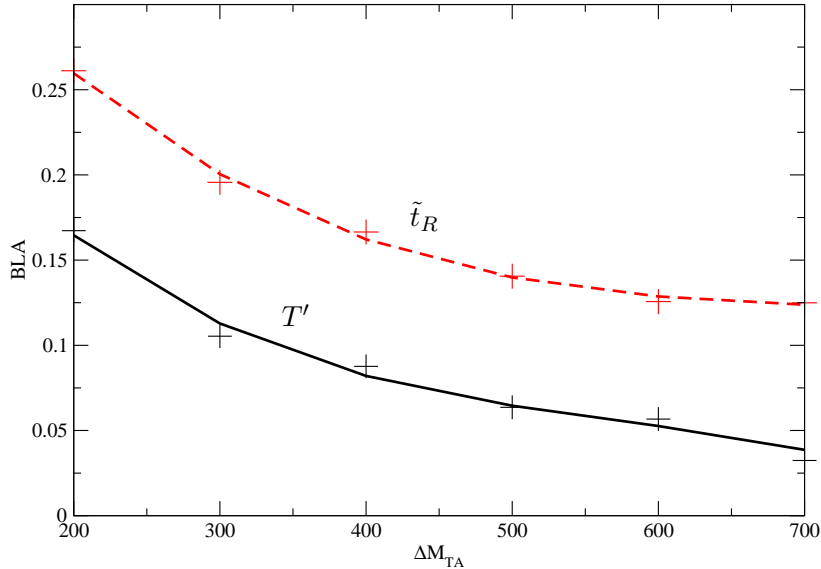}
\caption{Beamline asymmetry (BLA) as a function of the mass
difference between the top partner $T$ and the lightest parity odd particle $A_0$ in
a Little Higgs model (solid line) and SUSY (dashed line).  
Generator-level momenta
are used for illustration, since events cannot be fully reconstructed.}
    \label{bla}
\end{center}
\end{figure}

Therefore, we have to rely on subtle kinematical differences. As a
first example,  based on the fact that a scalar would be 
lighter than a fermion with a given production cross section, we could try to look for
some indication that the scalar is more boosted, using a parameter like the 
beamline asymmetry proposed in Ref.~\cite{tprime-study-1}
\be
\mbox{BLA} = \frac{N^z_{+} - N^z_{-}}{N^z_{+}+N^z_{-}},  
\ee
where $N^z_{+}$ ($N^{z}_{-}$) is the number of events with $p_{t_1}^z
p_{t_2}^z$$>$ ($<$) 0, for the momenta of the top quarks
$p_{t_1}$ and $p_{t_2}$. 
The beam-line asymmetry  for a $\tilde{t}_R$ in SUSY and
a $T'$ in a Little Higgs model with the same production cross section 
is presented in Fig.~\ref{bla}.  Note that we 
have not reconstructed the tops but simply used the known top momentum to 
illustrate the difference between the two cases.  
\psfrag{T}{$T'$}
\psfrag{stop}{$\tilde{t}_R$}
We remark that such a variable is also
subject to several potential sources of error. First of all,  the
average boost will be sensitive to the parton distribution functions, 
which typically give rise to 
an error of $5-10 \%$. More importantly, we must take into
account the effect of the SM backgrounds \cite{tprime-study-1}. Since
the asymmetry for scalars and fermions are distinguishable at the 
$10 \%$ level, we typically need an accuracy of  $S/\sqrt{S+B}
\sim 10$ in order to have a statistically significant effect. From the
results of Ref.~\cite{tprime-study-1}, as well as this study, we see that such a
high significance could be challenging for $m_{T} \sim 1$
TeV. Moreover, notice that the asymmetry decreases as the mass splitting
increases, since for small mass splitting the top quarks produced are closer to 
the direction of $T$ due to
phase-space limitations. This will further limit the utility of such
variables since, as we emphasized in this study, we will certainly have
more statistics for the signal with larger mass splittings.  
We also note that
constructing an analog of this observable in the semi-leptonic channel,
where some information about the top momentum will be inevitably lost, is
less straightforward. One may need to rely on the partially reconstructed top
momentum $p_b+p_{\ell}$ in the leptonic decay. As a result, we expect the mass information
contained in the beamline asymmetry variable to be less prominent. 

Using information about the production angular distributions to 
distinguish between fermionic and bosonic top partners will be challenging as well 
due to the presence of the additional missing neutrino, which makes the 
top quark reconstruction more difficult.  Obtaining clues from 
the directions of the visible decay products is possible, but since these are
also significantly affected by other factors such as the boost of the
top partner, the phase space, and the chirality of its coupling to 
the top quark, we expect any information about
the original production matrix element to be washed
out. Even if we are able to obtain such information, 
relating it to the spin of the top partner
is not at all straightforward. Unlike processes such as $Z \rightarrow
\ell^+ \ell^-$, pair production of top partners is not dominated by a
single s-channel process. Therefore, several partial waves may
contribute, making a distinction between fermion and boson potentially
challenging. 

We now comment on spin-determination strategies using invariant mass
distributions of objects in long decay chains
\cite{spin}. Consider a decay chain $X \rightarrow a +
Y^{(*)}(\rightarrow b+ Z) $, where $a$, $b$ are observable SM particles
and $X$, $Y$ and $Z$ are new particles beyond the SM, with $Y$ either
on or off-shell.  In this case, the invariant 
mass distribution $m_{ab}$ carries information about the spin 
of the {\it intermediate}
particle $Y$. In particular, it is a polynomial of order $2
J_{Y}$. Such a strategy will not be applicable in the simple 
decay $T \rightarrow t + A^0$ of our current interest, since $T$ is in the initial state. If
instead of this minimal setup, there is an extended top-partner sector,
with the possibility of it decaying through a longer decay
chain, we could in principle extract more information. For example, if
the top partner decays via $T \rightarrow b+W_{H}(\rightarrow W +
A^0 )$, the spin of $W$-partner $W_H$ could in principle be
measured by studing the correlation between $b$ and $W$. Then, assuming
Lorentz invariant and renormalizable interactions,  we could infer  the 
spin of the top partner. Considering this to be a plausible scenario, 
it seems worthwhile  carrying out a more detailed study and quantifying 
the observational feasibility for  this channel. 

\section{Summary and Conclusions}

We have studied methods of extracting the new physics signal in the $t \bar{t}
+ \not{\! \! E}_T$  
final state, with the tops decaying semi-leptonically.
We have developed kinematical variables 
that optimally separate the signal from the SM backgrounds.
The leading background is $t\bar t$ production from QCD. The other two large
backgrounds are $Wb\bar b jj$ and $t\bar tZ$. 
We found the following kinematical variables useful:

\begin{itemize}
\item A cut on missing transverse energy can effectively 
remove the $t\bar tZ$;
\item Looking inside the mass windows $m_{jj}\approx M_W$ and
 $m_{bjj}\approx m_t$ for the hadronically-decaying top 
will suppress $Wb\bar b jj$; 
\item The back-to-back nature of $t\bar t$ (or realistically
  $\phi_{t-b\ell}$) can be used to select against backgrounds;
\item The reconstruction of the transverse mass $M_T(W)$ can effectively single 
out  background events with a neutrino as the only missing particle;
\item Extending the definition of the reconstructed top mass $m_t^r$ 
in the leptonic mode
to include unphysical values can effectively separate out the large 
$t\bar t$ background.

\end{itemize}
In all the cases, if the mass difference is small $\dm\sim 200-300$ GeV, then
the signal kinematics, which are very much like those for $t\bar t$ production,
are very difficult to observe above the backgrounds.
The signal observability is summarized in Figs.~\ref{fig:MTDM} and
\ref{fig:MSTDM}. 

We have also commmented on  possible ways to distinguish between two typical
examples of such new physics: a $\tilde{t}_R$ in SUSY from a $T'$ in models 
with a
fermionic top partner. Such methods include using the difference in
the production rate, measuring angular correlations, and possibly
exploring differences in the $T'$ and $\tilde{t}_R$ couplings. We concluded
that none of these methods provide an easy determination of the
spin of the top partner in this particular channel, with only the minimal 
decay pattern assumed
in this study we are likely to have to explore more subtle kinematical
distributions, either to obtain a handle on the absolute mass scale
or precisely measure the production and decay matrix elements, or both
at the same time. We also note that if the new physics including the top
partner is beyond the minimal framework that we have focused on, this
generically makes both the detection and measurement of the properties of the
top partner easier. It would be interesting to explore some of these
possibilities in some detail.

\section{Acknowledgement}
We are grateful for discussions with Nima Arkani-Hamed, Robin
Erbacher, Henry Frisch, Kyoungchul Kong,   
Partick Meade and Mihoko Nojiri. We would also like to thank the Aspen Center
for Physics where part of the work was completed, for its 
hospitality. 
The work of T.H. was supported in part by the US DOE under contract No.~DE-FG02-95ER40896 
and in part by the Wisconsin Alumni Research Foundation.
The work of L.W.  is supported by the National Science Foundation
under Grant No.~0243680 and the Department of Energy under grant
No.~DE-FG02-90ER40542. Fermilab is operated by Fermi Research Alliance, 
LLC under Contract No. DE-AC02-07CH11359 with the United States Department of Energy.
The part of this research carried out at KITP was supported by the National
Science Foundation under Grant No. PHY05-51164.
Any opinions, findings, and conclusions or recommendations expressed
in this material are those of the authorss and do not necessarily
reflect the views of  the National Science Foundation.

\appendix

\section{Definitions of Transverse Variables}\label{sec:def_meff}

In general, a transverse mass variable can be formally defined
by projecting the momenta to the plane perpendicular to the
beam direction
\bea
\label{eq:mt}
M^2_T &=& (\sum_i E_{iT} )^2 - (\sum_i{\vec p}_{iT})^2,\\
E_{iT} &=& \sqrt{m_i^2 + {\vec p}_{iT}^2} ,
\eea
where $E_{T} = |\vec p_T|$ for a massless particle.\footnote{Experimentally, the transverse
energy is the quantity measured by the calorimeters; 
while the transverse momentum is determined by the  charge tracking system.} 
If there are unobservable particles, such as neutrinos or
new stable neutral particles, then the missing transverse energy (or 
momentum) is defined by
\be
\vec{\etmiss} = \vec{\ptmiss} = - \sum_o{\vec p}_{o T}.
\label{eq:miss}
\ee
where ${\vec p}_{o T}$ are the transverse momenta of the visible particles
in the event. Two remarks are in order here. First, the mass information for a
missing particle 
is lost in the above definition. Second,
if we consider the complete system of an event as in Eq.~(\ref{eq:mt}), 
then the second term vanishes as a consequence of Eq.~(\ref{eq:miss}).
This leads to a simple relation 
\begin{equation}
M_T = \sum_i E_{i T} .
\end{equation}
In its simplest form, one can just sum over the transverse momentum of each
individual observable (massless) particles to obtain the so-called ``effective
transverse mass", 
\begin{equation}
M^{\rm eff}_T = \sum_o E_{oT}  + \etmiss 
\end{equation}
Alternatively, one may consider the ``cluster transverse mass"
\begin{equation}
M^{c}_T = \sqrt{m_c^2 + {\vec p}^2_{c\,T}} + \etmiss ,
\end{equation}
where $m_c^2 = ( \sum_o p_o )^2$ and ${\vec p}_{c\,T} =  -\vec\etmiss$.   

\section{Top Reconstruction in Semileptonic Mode}
\label{top-recon}
 
First consider the QCD $t\bar t$ production, with one top quark decaying
hadronically and the other leptonically. 
Identifying the observed missing transverse momentum to be from a missing
neutrino, the conservation of transverse momentum gives
\begin{equation}
p_{\nu_T} = -\sum_{\rm visible}p_T =  \ptmiss
\end{equation}
We also use mass-shell conditions for the neutrino and the
leptonically-decaying $W$.
\begin{eqnarray}
&& p_\nu^2 = 0\label{equ:numassshell}\\
&& (p_\nu+p_l)^2 = 2 p_\nu \cdot p_l = M_W^2\label{equ:wmassshell}
\end{eqnarray}
We can then solve for $p_{\nu_L}$ to obtain, subject to the two-fold ambiguity, 
\begin{equation}
p_{\nu_L} = \frac{p_{l_L}(M_W^2-2\ptmiss
  .p_{l_T})\pm\sqrt{\Delta}}{2(E_l^2-p_{l_L}^2)} 
\end{equation}
where $\Delta$ is defined as
\begin{equation}
\Delta =E_l^2\left[M_W^4-4 M_W^2 \ptmiss
  \cdot p_{l_T}+ 4 (p_{l_T} \cdot  \ptmiss)^2-4(E_l^2-p_{l_L}^2)\ptmiss^2\right]  
\end{equation}
The correct solution for $p_\nu$ can be picked by enforcing that the
leptons plus $b$ reconstruct the top quark.  Grouping the wrong $b$-quark
with the leptons can be avoided by enforcing that the
hadronic jets also reconstruct the top quark.  In other words our solution
should minimize the following quantity, where $m_i$ stands for the
invariant mass of the reconstructed top quark on the $i$th fermionic leg:
\begin{equation}
\sum_{i=\rm l,\;h}\left|m_i^2-m_t^2\right|
\end{equation}

If we apply the same reconstruction procedure to the top quarks in
$T\overline{T}$ production, since we are overlooking the fact that
some of the missing energy in the event comes from the $A^0$s, rather
than the neutrino, there is no reason why our solution should
correspond to anything physical.  In fact, in general $\Delta$ might
even be a negative quantity, and we need a prescription to deal with
this case.  

We thus propose to generalize $p_\nu$ and $m_l^2$ to be complex quantities.  
When $\Delta <0$, we define
\bea
p_{\nu_L}&=&R_{\nu_L} \pm i I_{\nu_L}, \\
E_\nu &=& R_{\nu_E} \pm i I_{\nu_E} = R_{\nu_E} \pm i I_{\nu_L} \frac{p_{l_L}}{E_l} ,
\eea
where $R,\ I$'s are given in terms of the known quantities $p_l, \ptmiss, M_W$
as before. For definitiveness, we keep the $+$ sign for the solutions.
 We then reconstruct the complex top-quark mass
\bea
(m^r_t)^2 &=& (p_l+p_\nu+p_b)^2 = (E_l+E_b+R_{\nu_E} +  i I_{\nu_E})^2 \\
&-& (p_{lT}+p_{bT} + \ptmiss)^2 - (p_{lL}+p_{bL} +R_{\nu_L} +  i I_{\nu_L})^2.
\eea
The size of the imaginary part is a good measure of how ``far away"
from SM $\tt$ production the event is.

\end{document}